\documentclass[preprint]{aa} 
 \usepackage{lscape}
\usepackage{natbib}
\bibpunct{(}{)}{;}{a}{}{,} 
\usepackage{graphicx}
\usepackage[varg]{txfonts}

\def\asec{\ifmmode ^{\prime\prime}\else$^{\prime\prime}$\fi}

\def\it{\sl}
\def\degs{\ifmmode ^{\circ}\else$^{\circ}$\fi}
\def\amin{\ifmmode ^{\prime}\else$^{\prime}$\fi}
\def\asec{\ifmmode ^{\prime\prime}\else$^{\prime\prime}$\fi}

\def\obj{$\pi$ Aqr}
\def\degs{\ifmmode ^{\circ}\else$^{\circ}$\fi}
\def\amin{\ifmmode ^{\prime}\else$^{\prime}$\fi}

\def\eqalign#1{\null\,\vcenter{\openup1\jot \m@th
   \ialign{\strut\hfil$\displaystyle{##}$&$\displaystyle{{}##}$\hfil
   \crcr#1\crcr}}\,}
\begin{document}
   \title{Doppler tomography of the circumstellar disk of $\pi$~Aquarii.}
   \titlerunning{Doppler tomography of the circumstellar disk of $\pi$~Aqr.}
   \authorrunning{S. Zharikov, A. Miroshnichenko, et al.}
\author{S.~V.~Zharikov\inst{1},
A.~S.~Miroshnichenko\inst{2}, E.~Pollmann\inst{3},
S.~Danford\inst{2}, K.~S.~Bjorkman\inst{4}, N.~D.~Morrison\inst{4},
A.~Favaro\inst{5}, J.~Guarro Fl\'o\inst{6}, J.~N.~Terry\inst{7},
V.~Desnoux\inst{8}, T.~Garrel\inst{9}, G.~Martineau\inst{10},
Y.~Buchet\inst{10}, S.~Ubaud\inst{11}, B.~Mauclaire\inst{12},
H.~Kalbermatten\inst{13}, C.~Buil\inst{14}, C.~J.~Sawicki\inst{15},
T.~Blank\inst{16}, O.~Garde\inst{17} }

\institute{
Observatorio Astronomico Nacional, Instituto de
Astronomia, Universidad Nacional Autonoma de Mexico, Ensenada, BC,
Mexico \email{zhar@astrosen.unam.mx}
\and University of North Carolina at Greensboro, Greensboro, NC 27402, U.S.A.
\email{a$\_$mirosh@uncg.edu}
\and Emil-Nolde-Str. 12, 51375, Leverkusen, Germany
\and Ritter Observatory, University of Toledo, Toledo, OH 43606,
U.S.A
\and 19 Boulevard Carnot, 21000 Dijon, France
\and Balmes, 2, 08784, Piera (Barcelona), Spain
\and 6 rue Virgile, 42100, Saint-Etienne, France
\and ARAS, Astronomical Ring for Access to Spectroscopy,
http://www.astrosurf.com/aras/, France
\and Observatoire de Juvignac, 19 avenue du Hameau du Golf 34990,
Juvignac, France
\and SAPP, CSC des Trois Cit\'{E}s, Le Clos Gaultier, 86000,
Poitiers, France
\and 16 Calade, St.~Roch, 06410, Biot, France
\and Observatoire du Val de l'Arc, route de Peynier 13530, Trets,
France
\and Ebnetstrasse 12, CH-Bitsch, Switzerland
\and Castanet Tolosan Observatory, 6 place Cl\'emence Isaure 31320
Castanet Tolosan, France
\and Alpha Observatory, Alpine, Texas 79830, U.S.A.
\and Dorfstrasse 3f, 8603, Schwerzenbach, Switzerland
\and Observatoire de la Tourbi\'ere, 38690, Chabons, France}

   \date{Received - ; accepted -}

  \abstract
  {}
   {The work is aimed at a study of the circumstellar disk of the bright classical binary Be star $\pi$~Aqr. }
   {
   We analysed variations of a double-peaked profile of the $H_\alpha$ emission line in the spectrum of $\pi$
   Aqr that was observed in many phases during  $\sim$ 40 orbital cycles in 2004--2013. We applied the Discrete
   Fourier Transform (DFT) method to search for periodicity in the peak intensity (V/R) ratio. Doppler tomography
   was used to study the structure of the disk around the primary.
    }
   {The dominant frequency in the power spectrum of the $H_\alpha$ V/R ratio is 0.011873 day$^{-1}$ that correspond
    to a period of 84.2(2) days and is in agreement with the earlier determined orbital period of the system,
    $P_{orb}=84.1$ days. The V/R ratio shows a sinusoidal variation phase-locked with the orbital period.
    Doppler maps of all our spectra show a non-uniform structure of the disk around the primary: a ring with the
    inner and outer radii at $V_{in}\approx 450 $ km\,s$^{-1}$ and $V_{out} \approx  200$ km\,s$^{-1}$, respectively,
    along with an extended stable region (spot) at $V_x \approx$ 225 km\,s$^{-1}$ and $V_y\approx$ 100 km\,s$^{-1}$.
    The disk radius of $\approx 65 R_\odot = 0.33$ AU was estimated assuming Keplerian motion of a particle
    on a circular orbit at the disk outer edge.
}
   {}

 \maketitle
%

\section{Introduction}

Classical Be stars are fast-rotating non-supergiant B-type stars
with Balmer emission lines in their spectra.  The emission lines
are recombination lines that occur in a geometrically
thin, equatorial, 
circumstellar disk, according to a model first proposed by
\citet{1931ApJ....73...94S}. The model has been modified and further
developed by many authors and recently confirmed by direct
interferometric observations
\citep{1997ApJ...479..477Q,2007ApJ...654..527G,2009A&A...504..915C,2013AJ....145..141G}.
The disk may entirely disappear and re-appear unpredictably.  The
intensity of the emission lines varies on a time scale of days to
decades. It is widely accepted that the origin of the variability is
caused by processes occurring in the circumstellar disk and
non-radial pulsations of the B-star photosphere
\citep{2003PASP..115.1153P}. Typically the dominant feature in the
Be star spectra is an asymmetric double-peaked $H_\alpha$ emission
line.

 Many  Be stars exhibit variations in the ratio of the blue (violet)
and red emission peaks (V/R variations) of the Balmer emission lines
on a time scale of a few years. 
V/R variations may be due to  the evolution of a
one-armed spiral density pattern and/or binary effects
\citep{1983PASJ...35..249K,1991PASJ...43...75O,
1997A&A...318..548O}. The main observational properties of the V/R
variations and theoretical suggestions were summarized in
\citet{1997A&A...326.1167M}. The most complete compilation of Be
stars showing the V/R variations contains 62 objects
\citep{1997A&A...318..548O}. Eleven of them were known to be
binaries at the time. The time scale of the V/R variations is
typically much longer than rotational periods of the star or the
disk and does not correlate with orbital periods of the binaries.
However, some binary systems, $\varphi$ Per
\citep{1995A&A...304..235B,1998ApJ...493..440G}, V696 Mon
\citep{1972PASP...84..334P}, 4 Her \citep{1976BAICz..27...47H,
1997A&A...328..551K}, V744 Her \citep{1982A&A...115..138D}, $\kappa$
Dra \citep{1991BAICz..42...39J}, $\epsilon$ Cap
\citep{1999A&A...348..831R} and FY CMa \citep{2004A&A...427..307R}
present an exception from the rule.  For example, in the case of 4
Her, the V/R variations are orbital phase-locked and coherent over
more than 80 cycles \citep{2007ASPC..361..274S}.


\obj\ (HR 8539, HD 212571) is a bright, rapidly rotating ($\upsilon
\sin i \approx$ 300km\,s$^{-1}$) classical Be star with a variable
mass loss.
Analyzing its $H_\alpha$ line profiles and photospheric absorption
during a diskless phase in 1996--2001, \citet{2002ApJ...573..812B}
suggested that \obj\ is a binary system with an orbital period of
$P_{orb}=84.1$ days. The system consists of two stars with masses
$M_1\sin^3 i = 12.4 M_\odot$ and $M_2\sin^3 i = 2.0 M_\odot$. The
orbit is viewed at an inclination angle of $50^{\circ} -
75^{\circ}$. The components mass ratio and separation are
$M_2/M_1=0.16$ and $a = 0.96\sin^{-1} i$ AU, respectively. Using the
evolutionary tracks from \citet{1993A&AS..102..339S}, the effective
temperature T$_{\mathrm eff}^1=24000\pm1000$ of the primary component,
and its luminosity $\log (L_{bol}/L_\odot) = 4.1\pm0.3$,
\citep{2002ApJ...573..812B} estimated the primary's mass to be  $M_1
= 11\pm 1.5M\odot$. We further address this issue in
Sect.\,\ref{parameters}.

Variability of the Balmer line profiles of \obj, which appeared
double-peaked most of the time, has been reported by
\citet{1962ApJS....7...65M} and \citet{2012IBVS.6023....1P}.
\citet{1962ApJS....7...65M} observed strong V/R variations ranging
from 0.5 to 4.0, and several periods of absence of bright emission
lines (1936--1937, 1944--1945, 1950). Based on the \obj\ spectra
obtained between October 2004 and August 2011 together with the
available spectra of the data base
BeSS\footnote{http:/basebe.obspm.fr/basebe},
\citet{2012IBVS.6023....1P} found a dominant frequency in the power
spectrum of the V/R ratio that corresponded to a period of
$83.8\pm0.8$ days. This value coincides within the errors with the
above mentioned orbital period of the system.

 In this paper we attempt a new period analysis of the V/R variations
in the $H_\alpha$ line profile of $\pi$ Aqr based on 
spectroscopic observations 
obtained over $\sim$40 orbital cycles of the system in
2004--2013.
The disk structure of $\pi$ Aqr was probed using the Doppler
tomography method.  In Sect.\ref{obsred} we describe our
observations, in Sect.\ref{V/R} period analysis of the $H_\alpha$
line profile variation is presented. The Doppler tomography of the
system and modelling of the circumstellar disk are presented in
Sect.\ref{Dop}, and the results and conclusion are presented and
discussed in Sect.~\ref{discus}.

\begin{figure}[t]
\setlength{\unitlength}{1mm}
\resizebox{12cm}{!}{
\begin{picture}(120,60)(0,0)
\put (0,0) {\includegraphics[width=8.5 cm,clip]{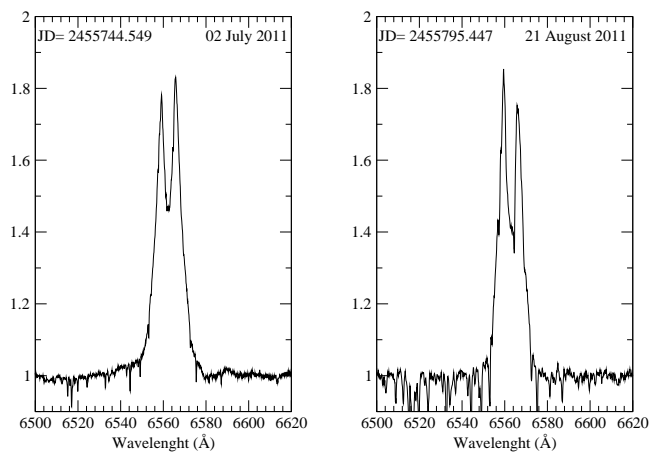}}
\end{picture}}
  \caption{V and R dominated examples of \obj\ $H_\alpha$ profiles. Epoch of observations are shown.}
 \label{fig1}
\end{figure}

\begin{figure}[ht]
\setlength{\unitlength}{1mm}
\resizebox{12cm}{!}{
\begin{picture}(120,47)(0,0)
\put (0,0) {\includegraphics[width=8.5 cm,clip]{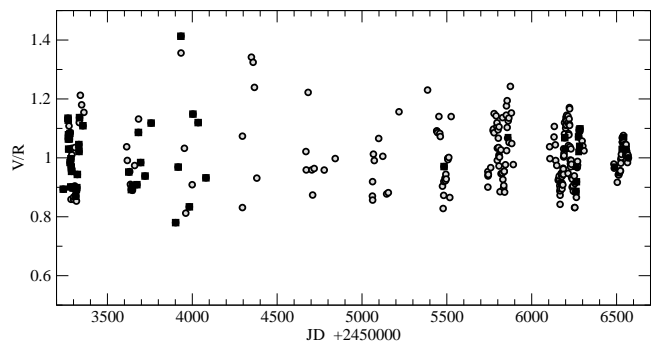}}
\end{picture}}
  \caption{Variation of V/R ratio of the $H_\alpha$ during reported observation.The   amateur  and professional data are marked by  grey circles and black squares, respectively.    }
 \label{fig2}
\end{figure}

\begin{figure}[ht]
\setlength{\unitlength}{1mm}
\resizebox{12cm}{!}{
\begin{picture}(120,65)(0,0)
\put (0,0) {\includegraphics[width=8.5 cm,clip]{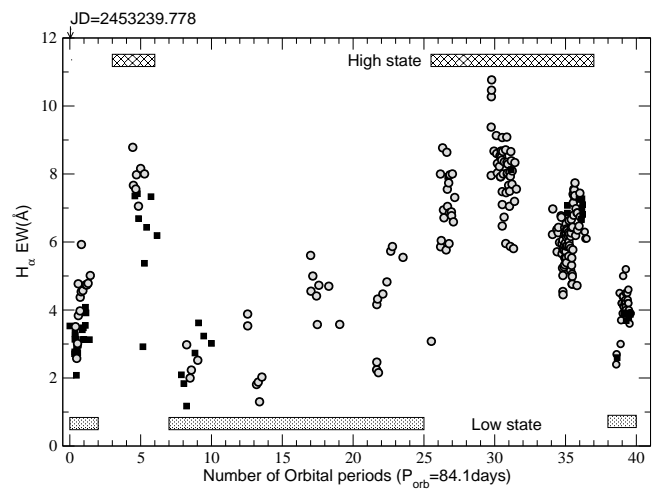}}
\end{picture}}
  \caption{The equivalent width of $H_\alpha$ emission line  vs. of the orbital cycles of the system. The epoch of beginning of  observations is shown too. The notes and bands in figure  mark values of the $H_\alpha$ equivalent widths selected for "high" and "low" state. Symbols are the same as in Fig.~\ref{fig2}. }
 \label{fig3}
\end{figure}

\section{Observations and data reduction}
\label{obsred}


 The object is continuously observed by both professionals and
amateurs since its binarity was revealed. For this study we used
spectra obtained by both communities. In particular, 36 spectra were
obtained in 2004--2006 with the 1.0\,m telescope of the Ritter
Observatory of the University of Toledo (Toledo, OH, USA) using a
fiber-fed \'echelle spectrograph with a Wright Instruments Ltd. CCD
camera. The spectra consisted of nine non-overlapping orders $\sim$
70 \AA\ in the range 5285--6597 \AA\ with a spectral resolving power
$R \simeq$ 26000.  Twelve spectra were obtained between October
2011--October 2013 at the 0.81\,m telescope of the Three College
Observatory (near Greensboro, NC, USA) using a fiber-fed \'echelle
spectrograph manufactured by Shelyak Instruments with a SBIG
ST--7XMEI  (2011--2012) or an ATIK--460EX (2013) CCD camera.
The spectra cover a range 4600--7200 \AA\ with a $R \simeq$ 10000.
These data were reduced with IRAF{\footnote {IRAF is distributed by
the National Optical Astronomy Observatories, which are operated by
the Association of Universities for Research in Astronomy, Inc.,
under contract with the National Science Foundation.}.

The contribution of the amateur community to the campaign involved
17 observers from Germany, France, Spain, Mexico, Switzerland, and
the USA. They used 0.2\,m to 0.4\,m telescopes with  a long-slit (in
most cases) and \'echelle spectrographs with a range of $R$ =
10000--22000. In total,  238 spectra were obtained between
September 2004 and October 2013. Data reduction was performed using
MaxIm-DL 3.06 (Diffraction Limited, Sehgal Corporation) for
Pollmann's data, while data from other amateurs were reduced with
software packages developed for amateur spectrographs, such as
SpcAudace\footnote{http://wsdiskovery.free.fr/spcaudace},
Audela\footnote{http://www.audela.org/dokuwiki/doku.php/en/start},
VSpec\footnote{http://www.astrosurf.com/vdesnoux/download.html}, and
IRIS3\footnote{http://www.astrosurf.com/buil/isis/isis$\_$en.htm}.
Spectral line parameters were measured with the spectral
classification software package
MK32\footnote{http://www.appstate.edu/$\sim$grayro/MK/MKbook.html}.
No systematic difference in the V/R ratios or the $H_\alpha$ line
equivalent widths (hereafter EW$_{H_{\alpha}}$) were found between
the professional and amateur data  (see Fig.~\ref{fig2}
\&~\ref{fig3}). We showed earlier that radial velocity data from
both communities also agree very well \citep{2013ApJ...766..119M}.
 The dates of observations, the source, and results of  the V/R and EW measurements are presented in Table~\ref{T:log}, which is available in the online version.
\onllongtab{
 \begin{longtable}{cccc}
 \caption{The dates of observations  of $\pi$ Aqr, the source,  and results of  V/R and EW measurements of H$_{\alpha}$ emission line.}\\
                       \hline \hline
		& & &  \\
HJD & V/R$_{H_{\alpha}}$ & EW$_{H_\alpha}$ & Reference \\ 
		& & &  \\ \hline
2453239.78&	0.89&	3.53&	1	\\	
2453267.69&	1.13&	2.71&	1	\\	
2453268.64&	1.13&	2.77&	1	\\	
2453269.67&	1.06&	3.14&	1	\\	
2453270.64&	1.08&	3.31&	1	\\	
2453271.72&	1.03&	3.25&	1	\\	
2453273.38&	1.04&	3.51&	2	\\	
2453275.60&	1.06&	3.14&	1	\\	
2453278.66&	1.08&	2.08&	1	\\	
2453279.66&	0.98&	2.95&	1	\\	
2453280.57&	1.02&	2.58&	2	\\	
2453283.67&	0.90&	2.80&	1	\\	
2453284.64&	1.00&	3.01&	1	\\	
2453285.41&	0.98&	3.01&	2	\\	
2453286.67&	0.97&	2.74&	1	\\	
2453288.67&	0.95&	2.96&	1	\\	
2453289.45&	1.00&	2.99&	2	\\	
2453289.59&	0.95&	4.77&	1	\\	
2453290.30&	1.00&	3.83&	2	\\	
2453297.54&	0.98&	3.93&	2	\\	
2453300.35&	0.96&	3.97&	2	\\	
2453301.41&	0.96&	4.38&	2	\\	
2453307.39&	0.97&	4.54&	2	\\	
2453308.31&	1.01&	5.93&	2	\\	
2453313.62&	0.87&	3.42&	1	\\	
2453316.57&	0.89&	3.13&	1	\\	
2453317.21&	0.97&	4.57&	2	\\	
2453319.58&	0.90&	3.48&	1	\\	
2453322.53&	0.94&	3.14&	1	\\	
2453331.57&	1.05&	3.55&	1	\\	
2453332.58&	1.02&	4.08&	1	\\	
2453334.19&	1.03&	4.74&	2	\\	
2453335.54&	1.14&	3.91&	1	\\	
2453340.18&	1.05&	4.74&	2	\\	
2453348.20&	1.04&	4.78&	2	\\	
2453355.52&	1.11&	3.13&	1	\\	
2453361.21&	1.04&	5.01&	2	\\	
2453613.35&	1.02&	8.78&	2	\\	
2453617.28&	0.99&	7.67&	2	\\	
2453625.66&	0.95&	7.35&	1	\\	
2453631.20&	0.98&	7.56&	2	\\	
2453635.29&	0.97&	7.98&	2	\\	
2453640.70&	0.89&	7.42&	1	\\	
2453647.31&	0.96&	6.69&	2	\\	
2453647.61&	0.90&	6.69&	1	\\	
2453660.29&	0.98&	8.16&	2	\\	
2453673.56&	0.91&	2.92&	1	\\	
2453682.61&	1.09&	5.37&	1	\\	
2453683.21&	1.05&	8.00&	2	\\	
2453695.59&	0.98&	6.43&	1	\\	
2453757.54&	1.12&	6.19&	1	\\	
2453901.84&	0.78&	2.09&	1	\\	
2453916.83&	0.97&	1.83&	1	\\	
2453932.83&	1.41&	1.17&	1	\\	
2453933.55&	1.08&	2.98&	2	\\	
2453953.56&	1.01&	2.00&	2	\\	
2453961.52&	0.96&	2.23&	2	\\	
2453982.70&	0.83&	2.73&	1	\\	
2453999.38&	0.98&	2.52&	2	\\	
2454003.65&	1.15&	3.62&	1	\\	
2454034.67&	1.12&	3.23&	1	\\	
2454080.49&	0.93&	3.02&	1	\\	
2454295.55&	0.96&	3.88&	2	\\	
2454295.58&	1.02&	3.53&	2	\\	
2454348.40&	1.05&	1.81&	2	\\	
2454358.44&	1.07&	1.88&	2	\\	
2454366.36&	1.03&	1.30&	2	\\	
2454380.35&	0.99&	2.03&	2	\\	
2454669.49&	1.01&	5.61&	2	\\	
2454671.47&	0.99&	4.55&	2	\\	
2454682.50&	1.03&	5.00&	2	\\	
2454704.41&	0.99&	4.41&	2	\\	
2454709.37&	0.97&	3.57&	2	\\	
2454718.42&	0.99&	4.73&	2	\\	
2454777.35&	0.99&	4.70&	2	\\	
2454842.23&	1.00&	3.58&	2	\\	
2455061.39&	0.97&	2.24&	2	\\	
2455062.44&	0.98&	4.16&	2	\\	
2455062.56&	0.97&	2.47&	2	\\	
2455067.39&	0.99&	4.32&	2	\\	
2455072.41&	1.00&	2.16&	2	\\	
2455098.38&	1.01&	4.47&	2	\\	
2455122.48&	1.01&	4.83&	2	\\	
2455145.30&	0.97&	5.73&	2	\\	
2455155.26&	0.95&	5.87&	2	\\	
2455218.26&	1.05&	5.55&	2	\\	
2455386.55&	1.07&	3.08&	2	\\	
2455440.39&	1.04&	8.00&	2	\\	
2455441.45&	1.03&	5.86&	2	\\	
2455445.37&	1.03&	6.04&	2	\\	
2455453.33&	1.05&	8.77&	2	\\	
2455458.44&	1.03&	6.94&	2	\\	
2455461.32&	1.03&	6.71&	2	\\	
2455474.31&	0.97&	5.77&	2	\\	
2455478.28&	0.92&	8.64&	2	\\	
2455481.28&	0.95&	7.55&	2	\\	
2455482.29&	0.97&	7.05&	2	\\	
2455489.46&	0.97&	7.74&	2	\\	
2455491.31&	0.97&	5.95&	2	\\	
2455494.29&	0.98&	6.86&	2	\\	
2455496.34&	0.96&	7.97&	2	\\	
2455505.42&	1.00&	6.89&	2	\\	
2455506.40&	0.99&	6.77&	2	\\	
2455511.46&	1.00&	8.00&	2	\\	
2455517.25&	0.95&	6.59&	2	\\	
2455525.35&	1.06&	7.31&	2	\\	
2455741.58&	0.97&	7.96&	2	\\	
2455741.65&	0.98&	9.38&	2	\\	
2455742.60&	0.97&	10.27&	2	\\	
2455743.66&	0.97&	10.46&	2	\\	
2455744.65&	0.97&	10.77&	2	\\	
2455758.49&	0.98&	8.67&	2	\\	
2455772.50&	1.03&	9.13&	2	\\	
2455774.55&	1.05&	8.60&	2	\\	
2455778.48&	1.04&	8.03&	2	\\	
2455778.53&	1.06&	8.31&	2	\\	
2455790.49&	1.06&	8.22&	2	\\	
2455794.46&	1.04&	8.54&	2	\\	
2455795.45&	1.04&	8.53&	2	\\	
2455797.39&	1.00&	7.91&	2	\\	
2455797.54&	1.02&	7.92&	2	\\	
2455801.46&	1.03&	8.68&	2	\\	
2455803.39&	1.02&	8.51&	2	\\	
2455803.46&	1.01&	 	&2		\\
2455804.49&	1.00&	8.41&	2	\\	
2455807.45&	1.05&	6.47&	2	\\	
2455807.45&	1.01&	7.48&	2	\\	
2455807.52&	0.99&	9.07&	2	\\	
2455810.35&	1.01&	8.66&	2	\\	
2455813.39&	0.98&	8.00&	2	\\	
2455818.41&	1.01&	6.73&	2	\\	
2455819.44&	0.99&	8.36&	2	\\	
2455827.53&	1.01&	8.71&	2	\\	
2455828.34&	1.05&	5.96&	2	\\	
2455830.37&	0.97&	7.45&	2	\\	
2455834.31&	0.95&	9.09&	2	\\	
2455835.37&	0.98&	8.34&	2	\\	
2455837.45&	0.98&	8.00&	2	\\	
2455838.36&	0.99&	8.03&	2	\\	
2455840.40&	0.96&	8.29&	2	\\	
2455845.36&	0.99&	7.67&	2	\\	
2455850.29&	1.06&	7.05&	2	\\	
2455851.26&	1.06&	7.49&	2	\\	
2455852.32&	1.01&	8.05&	2	\\	
2455855.28&	1.08&	5.87&	2	\\	
2455855.37&	1.05&	7.93&	2	\\	
2455858.39&	1.04&	8.66&	2	\\	
2455860.69&	1.07&	8.13&	3	\\	
2455861.37&	1.04&	8.39&	2	\\	
2455865.28&	1.07&	7.70&	2	\\	
2455866.52&	1.04&	8.09&	2	\\	
2455874.29&	1.10&	5.81&	2	\\	
2455880.36&	1.07&	7.19&	2	\\	
2455882.27&	1.02&	8.34&	2	\\	
2455891.39&	0.99&	7.56&	2	\\	
2456105.63&	1.01&	6.22&	2	\\	
2456107.62&	0.99&	6.97&	2	\\	
2456131.54&	1.04&	6.36&	2	\\	
2456133.45&	1.04&	5.71&	2	\\	
2456139.47&	1.02&	6.27&	2	\\	
2456141.50&	1.01&	6.43&	2	\\	
2456147.55&	0.99&	5.99&	2	\\	
2456156.52&	0.97&	6.77&	2	\\	
2456158.47&	0.97&	6.73&	2	\\	
2456162.43&	0.98&	5.96&	2	\\	
2456162.45&	0.97&	5.65&	2	\\	
2456165.44&	0.97&	5.23&	2	\\	
2456167.37&	0.95&	5.78&	2	\\	
2456167.44&	0.95&	5.01&	2	\\	
2456167.46&	0.95&	4.54&	2	\\	
2456167.47&	0.94&	6.06&	2	\\	
2456168.41&	0.95&	4.45&	2	\\	
2456172.38&	0.96&	5.82&	2	\\	
2456174.39&	0.98&	5.27&	2	\\	
2456177.32&	0.96&	5.76&	2	\\	
2456177.40&	0.99&	5.90&	2	\\	
2456177.50&	0.98&	6.23&	2	\\	
2456178.37&	0.96&	6.22&	2	\\	
2456178.43&	0.96&	5.33&	2	\\	
2456178.47&	0.98&	6.42&	2	\\	
2456179.35&	0.96&	5.54&	2	\\	
2456180.45&	0.97&	5.53&	2	\\	
2456183.34&	0.96&	5.98&	2	\\	
2456185.32&	0.98&	5.72&	2	\\	
2456186.36&	0.99&	6.54&	2	\\	
2456186.42&	0.98&	6.23&	2	\\	
2456187.31&	0.98&	5.54&	2	\\	
2456188.35&	0.99&	5.39&	2	\\	
2456188.41&	1.00&	5.88&	2	\\	
2456190.35&	1.01&	6.13&	2	\\	
2456190.41&	1.00&	6.39&	2	\\	
2456193.30&	1.02&	6.35&	2	\\	
2456193.45&	1.04&	5.90&	2	\\	
2456194.64&	1.03&	6.85&	3	\\	
2456200.34&	1.04&	6.28&	2	\\	
2456200.61&	1.06&	6.16&	2	\\	
2456202.42&	1.05&	5.58&	2	\\	
2456203.45&	1.05&	6.00&	2	\\	
2456204.33&	1.05&	6.63&	2	\\	
2456204.40&	1.05&	5.87&	2	\\	
2456205.34&	1.06&	6.24&	2	\\	
2456206.29&	1.06&	5.89&	2	\\	
2456213.40&	1.05&	5.81&	2	\\	
2456213.46&	1.04&	5.16&	2	\\	
2456214.40&	1.05&	6.16&	2	\\	
2456216.39&	1.02&	6.27&	2	\\	
2456217.33&	1.04&	6.43&	2	\\	
2456217.38&	1.02&	6.66&	2	\\	
2456219.27&	1.03&	5.31&	2	\\	
2456220.37&	1.06&	5.53&	2	\\	
2456222.26&	1.05&	4.76&	2	\\	
2456222.34&	1.03&	5.05&	2	\\	
2456223.29&	1.05&	4.99&	2	\\	
2456223.43&	1.04&	4.76&	2	\\	
2456225.42&	1.01&	6.44&	2	\\	
2456226.35&	0.99&	5.77&	2	\\	
2456228.26&	0.99&	7.41&	2	\\	
2456229.37&	0.99&	7.15&	2	\\	
2456231.35&	1.00&	6.79&	2	\\	
2456232.35&	1.00&	7.55&	2	\\	
2456234.35&	1.00&	7.35&	2	\\	
2456235.24&	0.98&	7.53&	2	\\	
2456238.30&	0.95&	7.36&	2	\\	
2456239.30&	0.98&	7.74&	2	\\	
2456243.24&	0.96&	6.19&	2	\\	
2456245.27&	0.95&	6.98&	2	\\	
2456251.31&	0.97&	4.72&	2	\\	
2456252.28&	0.95&	6.86&	2	\\	
2456254.20&	0.96&	6.86&	2	\\	
2456254.32&	0.95&	6.35&	2	\\	
2456263.35&	0.96&	6.47&	2	\\	
2456263.48&	0.97&	7.26&	3	\\	
2456268.30&	0.99&	7.44&	2	\\	
2456270.34&	0.99&		&2		\\
2456270.35&	0.99&	6.74&	2	\\	
2456276.48&	1.02&	6.67&	3	\\	
2456280.48&	1.04&	7.00&	3	\\	
2456284.46&	1.04&	6.81&	3	\\	
2456297.23&	1.06&	6.30&	2	\\	
2456297.29&	1.03&		&2		\\
2456301.28&	1.04&	6.10&	2	\\	
2456308.25&	1.02&	6.10&	2	\\	
2456484.50&	0.98&	2.40&	2	\\	
2456486.56&	0.97&	2.70&	2	\\	
2456489.70&	0.97&	2.60&	3	\\	
2456504.42&	0.92&	4.50&	2	\\	
2456510.38&	0.94&	3.00&	2	\\	
2456513.50&	0.94&	3.70&	2	\\	
2456516.43&	0.95&	4.20&	2	\\	
2456518.38&	0.95&	4.20&	2	\\	
2456522.39&	0.96&	3.90&	2	\\	
2456522.44&	0.96&	4.40&	2	\\	
2456525.37&	1.00&	5.00&	2	\\	
2456525.44&	0.98&	4.40&	2	\\	
2456528.43&	1.00&	4.10&	2	\\	
2456528.44&	1.01&	4.10&	2	\\	
2456530.71&	1.03&	3.92&	3	\\	
2456533.39&	1.07&		&2		\\
2456534.43&	1.05&	4.20&	2	\\	
2456534.79&	1.05&	3.90&	2	\\	
2456536.76&	1.04&		&2		\\
2456538.42&	1.06&	4.30&	2	\\	
2456539.43&	1.07&	4.20&	2	\\	
2456539.45&	1.07&	4.20&	2	\\	
2456540.42&	1.08&	4.10&	2	\\	
2456541.35&	1.05&	5.20&	2	\\	
2456541.40&	1.04&	4.50&	2	\\	
2456541.44&	1.04&	4.00&	2	\\	
2456543.63&	1.07&	3.69&	3	\\	
2456549.75&	1.05&	4.50&	2	\\	
2456550.82&	1.02&		&2		\\
2456552.47&	1.05&	4.30&	2	\\	
2456554.29&	1.01&	4.20&	2	\\	
2456556.46&	1.03&	4.60&	2	\\	
2456556.47&	1.05&	4.20&	2	\\	
2456557.45&	1.02&		&2		\\
2456558.57&	1.03&	3.77&	3	\\	
2456559.41&	1.02&	4.00&	2	\\	
2456560.43&	1.01&		&2		\\
2456561.30&	1.01&		&2		\\
2456563.39&	1.01&	3.60&	2	\\	
2456564.69&	0.98&	3.90&	2	\\	
2456567.34&	1.01&	3.90&	2	\\	
2456567.58&	1.00&	3.87&	3	\\	
2456569.69&	1.00&	3.90&	2	\\	
2456570.62&	1.01&	3.80&	3	\\	
\hline

\label{T:log}
\end{longtable}
\tablefoot{ 1 - Ritter Observatory;
2 - amateur data;
3 -  Three College Observatory
}
}


\begin{figure}[t]
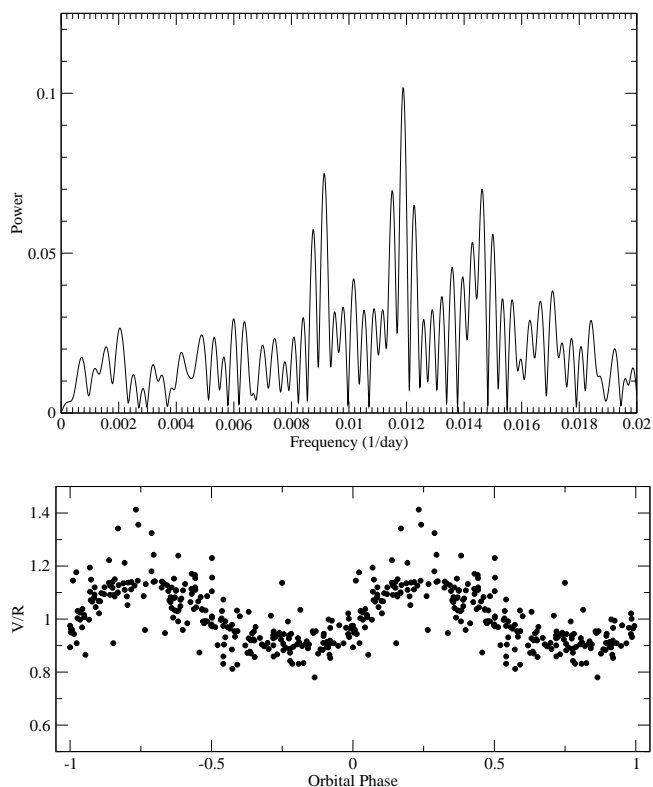

\setlength{\unitlength}{1mm}
\resizebox{12cm}{!}{
\begin{picture}(120,105)(0,0)
\put (0,45) {\includegraphics[width=8.5 cm,clip]{fig04a.eps}}
\put (0,0) {\includegraphics[width=8.5 cm,clip]{fig04b.eps}}
\end{picture}}
  \caption{Top panel: The power spectrum of V/R variation. Bottom panel: The V/R ratio folded on the orbital period of the system 84.1 day.
  $\phi_{orb} = 0.0$ corresponds to the inferior conjunction of the secondary.}
 \label{fig4}
\end{figure}

\begin{figure*}[ht]
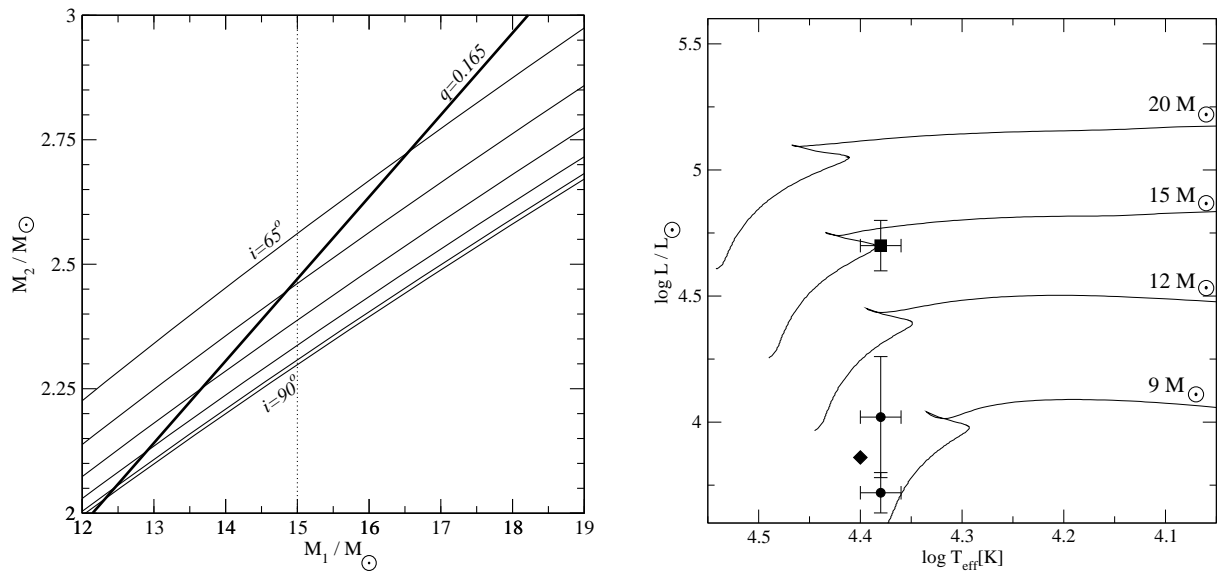

\setlength{\unitlength}{1mm}
\resizebox{12cm}{!}{
\begin{picture}(120,75)(0,0)
\put (95,0) {\includegraphics[width=7.5 cm,clip]{fig05a.eps}}
\put (10,0) {\includegraphics[width=7.75 cm,clip]{fig05b.eps}}
\end{picture}}
  \caption{  Left panel: M$_2$ vs M$_1$ relationships for different inclination $i$ of the system orbital
   plane obtained from the mass function  $f(M) = \frac{K_1^3 P_{orb}}{2 \pi G } = 0.041 M_\odot$.
   The inclination range is of 65$^{\circ}$ --- 90$^{\circ}$ with a step of 5$^{\circ}$ from top to bottom.
   The solid-line corresponds to a companions mass ratio of $q=0.165$. Main-sequence stars with $T_{\rm eff}=24000$K
   are located to the left from the dotted-line.
   Right panel: A Hertzsprung-Russell diagram with recent determinations of
  fundamental parameters of $\pi$~Aqr. Solid lines show evolutionary tracks of stars with
  rotation from \citet{2012A&A...537A.146E}. Initial masses are indicated by numbers at
  the corresponding track. Filled circles show the positions based on the luminosity
  calculated using the two {\it HIPPARCOS} distances (see text), the diamond shows data
  from \citet{2010AN....331..349H}, and the filled square shows parameters adopted here.}
 \label{fig5}
\end{figure*}

\section{Variability of the $H_\alpha$ line}\label{V/R}

As noted above, the $H_\alpha$ line profile of \obj\ is
double-peaked and strongly variable. Examples of the $H_\alpha$ line
profile for two epochs of observations are shown in
Figure~\ref{fig1}. We measured the V/R ratio  (defined as $V/R =
I_V/I_R$) in the $H_\alpha$ emission line using peak intensities of
the V and R components which were separately fitted with a single
gaussian. The values of the V/R ratio vary in the range of $0.8-1.4$
(Fig.~\ref{fig2}) and do not correlate with EW$_{H_{\alpha}}$
(Fig.~\ref{fig3}). The latter ranges between 1.0 and 11.0 \AA\ and
shows no periodicity. We refer to the spectra with a stronger
$H_\alpha$ line (EW$_{H_{\alpha}} \ge 4.8 \AA$) as to a ``high
state'' and to those with a weaker $H_\alpha$ line as to a ``low
state'' for the following analysis. The Discrete Fourier
Transform (DFT)
method\footnote{http://www.univie.ac.at/tops/Period04/} was used  to
search for periodicity in the V/R variation during our observations.
The dominant frequency in the power spectrum is 0.011873$\pm$0.00028
day$^{-1}$ corresponds to a period of $84.2\pm0.2$ days that is in
agreement with the system orbital period of $P_{orb}=84.1$ days
found by \citet{2002ApJ...573..812B}. The resulting power spectrum
and the V/R ratio folded with the dominant frequency are presented
in Figure~\ref{fig4}. Therefore, we conclude that the V/R variations
in $\pi$ Aqr are locked with the orbital period, at least on a time
scale of  $\approx 3400$ days or 40 orbital cycles. The shape
of the V/R variation curve is sinusoidal.

\begin{figure*}[ht]
\setlength{\unitlength}{1mm}
\resizebox{7.5cm}{!}{
\begin{picture}(70,100)(0,0)
\put (0,100){\includegraphics[width=9.5cm, bb = 145 185 460 750, angle=-90, clip=]{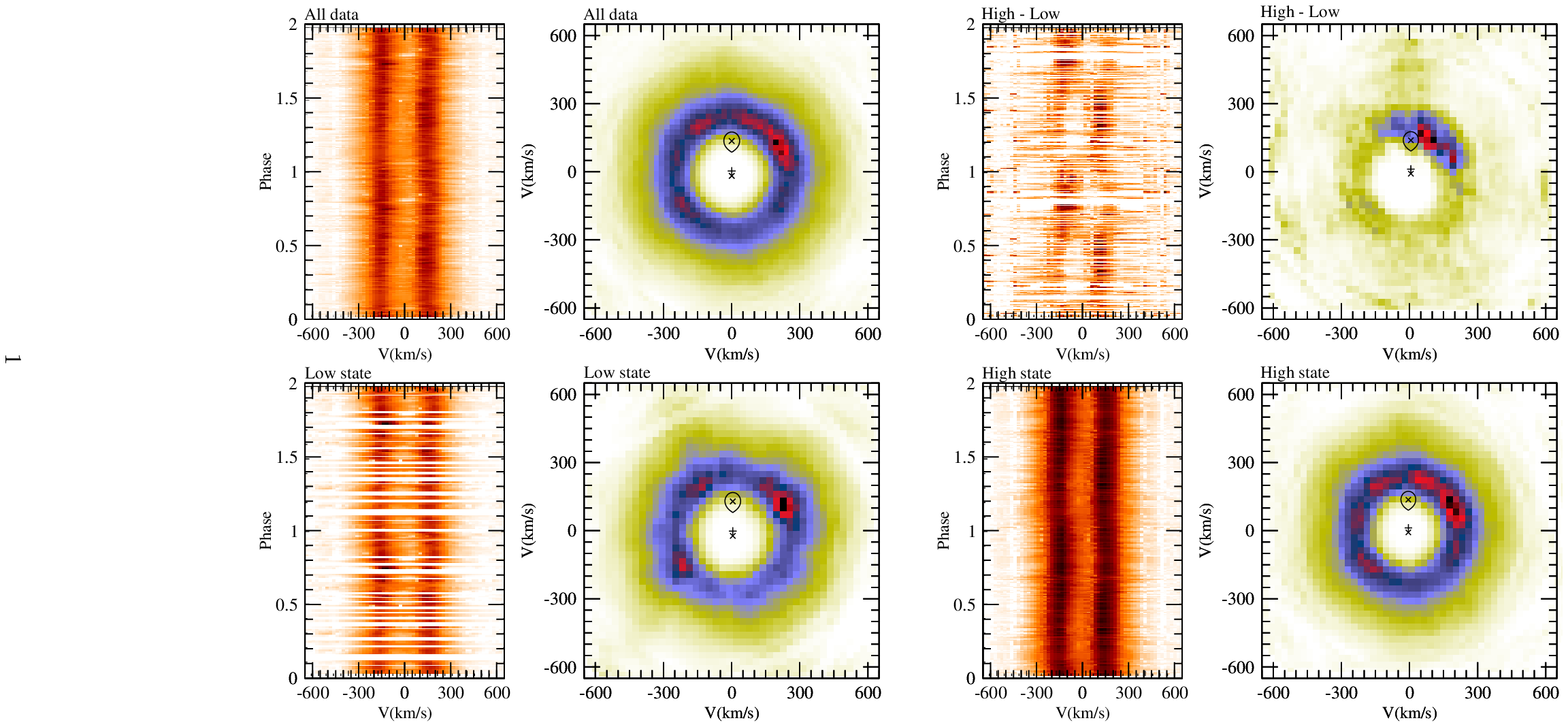}}
\end{picture}}
\caption{ The phased time series spectra around H$_\alpha$ line
 folded with the orbital period of the system and the corresponding Doppler maps are shown.
 The orbital period of $P_{orb} = 84.1$ days, the primary mass of $M_1 = 14.0 M_\odot$ ,
 and the mass ratio of q = 0.16 are used to overlay positions of the stellar components
 on the Doppler maps. The inclination angle $i = 70^{\circ}$ is arbitrarily chosen based
 on suggestions by \citet{2002ApJ...573..812B}.
 $\phi_{orb} = 0.0$ corresponds to the inferior conjunction of the secondary.}
\label{Dopmap}
\end{figure*}

\begin{figure*}[ht]
\setlength{\unitlength}{1mm}
\resizebox{7.5cm}{!}{
\begin{picture}(70,70)(0,0)
\put (10,-5){\includegraphics[width=7.5cm,bb= 90 60 450 750, angle=90,  clip=]{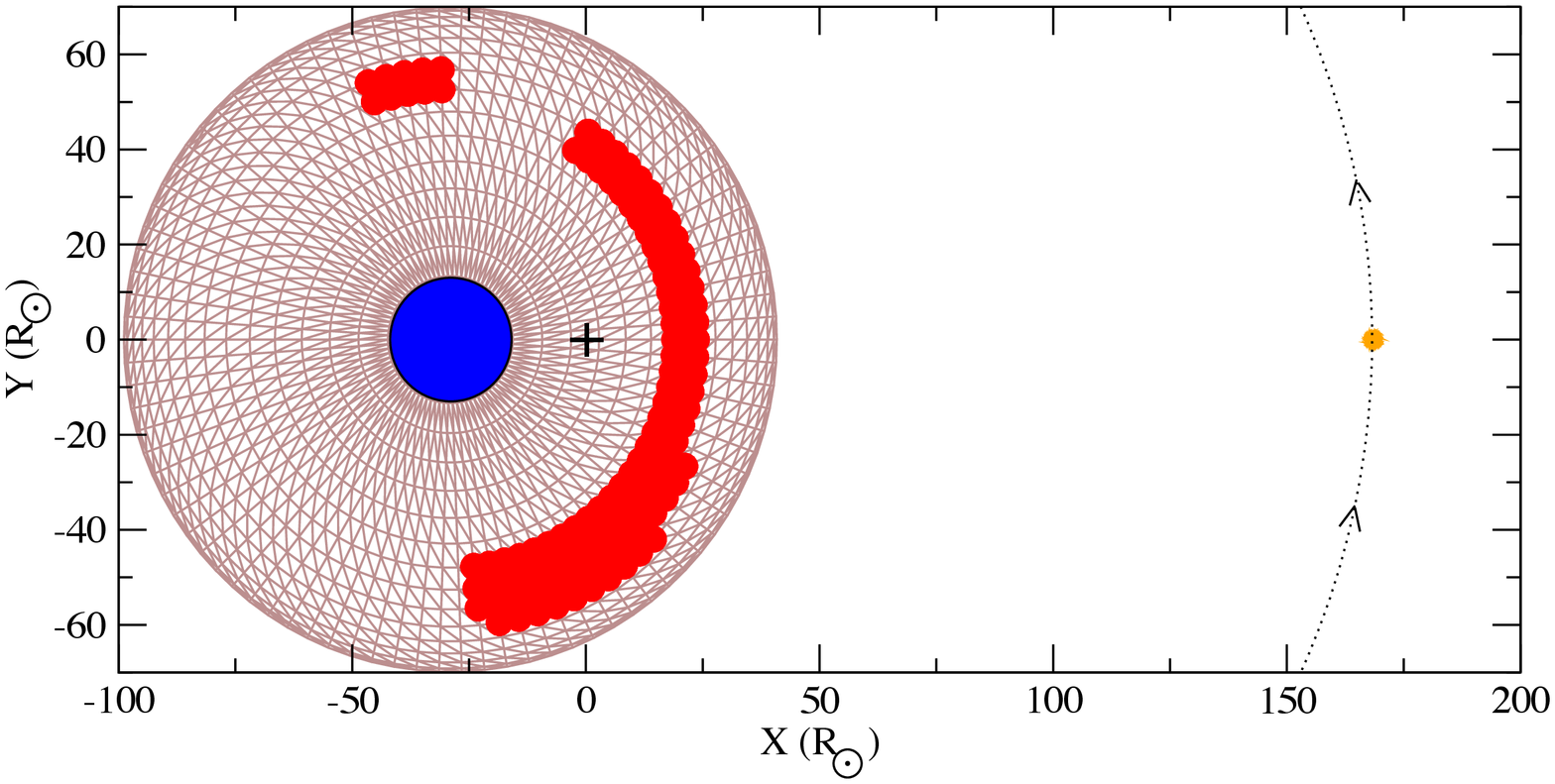}}
\end{picture}}
\caption{Our geometrical model of $\pi $ Aqr.  The blue and orange
filled circles mark the primary and the secondary, respectively. The
circumstellar disk is shown in brown colour. The red
regions in the disk are constructed on the basis of the Doppler
tomography and correspond  the excess of $H_\alpha$ emission comparison with the rest of the disk.
Scales of both axes are given in solar radii. Arrows show the
direction of the binary rotation, and the cross indicates the centre
of mass of the system. Parameters of the system are given in
Table~\ref{T:par}. } \label{Model}
\end{figure*}

\section{Doppler tomography and Disk model.}

\subsection{System parameters}\label{parameters}



There is an apparent discrepancy between the dynamical and the
 evolutionary mass of the primary component that was outlined.  This problem was not
 resolved by \citet{2002ApJ...573..812B}, therefore  we reanalyzed  below  system parameters of the $\pi$~Aqr binary.

We calculated an absolute magnitude of $M_V=-2.96^{+0.50}_{-0.58}$
mag using the following: system brightness during the diskless epoch
(1996--2001) of $V=4.85$ mag, an interstellar extinction of
$A_V=0.15$ mag, and a {\it HIPPARCOS} distance $340^{+105}_{-70}$pc
\citep{ESA1997}. A more recent distance $240^{+17}_{-15}$pc
\citep{2007A&A...474..653V} gives an absolute magnitude of
$M_V=-2.20^{+0.15}_{-0.16}$ mag. Applying a bolometric correction of
$BC_V=-2.36\pm0.10$ mag for $T_{\rm eff}= 24000\pm1000 K$
\citep{1997IAUS..189P..50M}, one gets a luminosity of $\log$
L/L$_{\odot}=4.02\pm0.24$ for the larger distance and $\log$
L/L$_{\odot}=3.72\pm0.08$ for the smaller one.  These
luminosities correspond to initial masses of 10.5 M$_{\odot}$ and
9.5 M$_{\odot}$, respectively, on the most recent evolutionary
tracks with rotation by \citet{2012A&A...537A.146E}.
Both these
 values are noticeably lower than the primary's dynamical mass $M_1\sin^3 i = 12.4
 M_{\odot}$ from \citet{2002ApJ...573..812B}. Since the radial
 velocity curves of both companions were well-established over the entire diskless
 period, the dynamical mass seems to be more reliable than the
 evolutionary mass.
  As seen in Fig.\ref{fig5} (left panel), the primary's mass ranges within
12.5---17.0~M$_{\odot}$ for the most probable orbital inclination of
$i=65-85^{\circ}$ \citep{2002ApJ...573..812B}. This mass range
requires a higher luminosity of the system (see Fig.\ref{fig5},
right panel).  Comparison with the evolutionary models by
\citet{2012A&A...537A.146E} also shows that the primary component
with $T_{\mathrm eff}^1= 24000 K$ in $\pi$ Aqr has evolved out of the
main-sequence if its mass exceeds $\sim$15 M$_{\odot}$.

  As Be stars are considered to be
 main-sequence objects \citep[e.g.,][]{2006A&A...451.1053F}, we
 therefore adopt a mass of 14.0$\pm$1.0 M$_{\odot}$ for the primary.
 This constrains its luminosity at $\log$ L/L$_{\odot}=4.7\pm0.1$,
 which in turn leads to $M_V=-4.64\pm0.25$ mag and a distance of
 740$\pm$90 pc. A separate study of stars near the object's line of
 sight (which is beyond the scope of this paper) is needed to verify this result.
 Nevertheless, the mass adjustment will only change the circumstellar disk scale,
 but not the qualitative results of our modelling.
 In  Table~\ref{T:par} we present a summary of adopted parameters of the $\pi$ Aqr binary system.

\begin{table}[t]
 \caption{Adopted parameters of $\pi$ Aqr.}
\begin{tabular}{ll}                                  \hline
Period                                &   84.1(1)~days          \\
M$_1$            &    14.0(1.0)$~M_\odot$       \\
$T_{\mathrm eff}^1$                         & 24000(1000)~K \\
$\log$~L/L$_\odot$                        & 4.7(1) \\
$R_1$$^*$                               & 13.0(1.4)$R_\odot$ \\
$\upsilon_c$                                          & 453km\,s$^{-1}$ \\
$M_V$$^*$                                & -4.64(25)       \\
$ q$  mass ratio                 &   $\geq$ 0.165                     \\
$i$ system inclination               &    65$^{\circ}$-85$^{\circ}$                 \\
M$_2$  & 2.31(16)$M_\odot$ \\
$a$ separation & 0.96 AU \\
Distance &  740(90) pc \\ \hline \\
\end{tabular} \\
\begin{list}{}
\item $^*$ based on L/L$_\odot$ and $T_{\mathrm eff}^1$
\item   numbers in brackets are $1\sigma$ uncertainties referring
to last significant digits quoted

\end{list}

\label{T:par}
\end{table}


\subsection{Doppler tomography}\label{Dop}
The $H_\alpha$ line profiles of $\pi$ Aqr were recorded at many
orbital phases of the system over $\sim$40 orbital cycles. Our
finding that the V/R variations are locked with the orbital period
suggests that the $H_\alpha$ line profile variability could be
caused by a complex structure in the circumstellar disk. We used
Doppler tomography \citep{1988MNRAS.235..269M} to study the
structure of the disk around primary in \obj.

The Doppler maps were built by combining time-resolved spectra using
the maximum entropy method as implemented by
\citet{1998astro.ph..6141S}\footnote{http://www.mpa-garching.mpg.de/$\sim$henk/pub/dopmap}.
Figure\,{\ref{Dopmap} shows 
phased time series spectra around the $H_\alpha$ line and
corresponding Doppler maps of all spectra (top raw, left panel),
``low'' (bottom raw, left panel) and ``high'' (bottom raw, right
panel) states, and a difference between each spectrum in a "high"
and the average spectrum in the ``low'' state (top raw, right
panel). The orbital period of P$_{\mathrm {orb}}=84.1$\,days, the
primary mass of M$_{1}=14.0$\,M$_\odot$, and the mass ratio $q=0.16$
(Table~\ref{T:par}) are used to overlay  positions of the stellar
components on the Doppler maps. The inclination angle $i=70^{\circ}$
is arbitrarily chosen based on the discussion in
Sect.\,\ref{parameters}. $\phi_{orb}$ = 0.0 corresponds to the
inferior conjunction of the secondary. The location of the main
components in the system, such as the position of the centre of
mass, the primary, the secondary, and the Roche lobe of the
secondary star, are indicated.

The Doppler maps of all spectra show a non-uniform structure of the
disk around the primary: a ring with the inner and outer radii at
$V_{in}\approx 450 $ km\,s$^{-1}$ and $V_{out} \approx  200$
km\,s$^{-1}$, respectively, together with an extended  stable region
(spot) at $V_x \approx$ 225 km\,s$^{-1}$ and $V_y\approx$ 100
km\,s$^{-1}$. We note that EW$_{H_\alpha}$ is a function of the
total flux and surface brightness distribution of the disk. The
"spot" is brighter and more extended in the ``High'' state compared
to the ``Low"'' state. However, the brightness of the extended
region is significantly lower than the total disk brightness and
corresponds to an S-wave that can be seen only in "High-Low" trailed
spectra. The disk radius of $\approx 65 R_\odot = 0.33$ AU was
estimated assuming Keplerian motion of a particle in a circular
orbit at the disk outer edge. Based on the results of the Doppler
tomography, a geometrical model of the $\pi$ Aqr system is presented
in Figure~\ref{Model}.


\section{Discussion and Conclusion}
\label{discus}


In this paper we attempted a new period analysis of the V/R
variations in the $H_\alpha$ line profile of $\pi$ Aqr and probe the
primary's circumstellar disk structure based on spectroscopic
observations obtained over multiple orbital cycles of the system.
The main conclusions of our analysis are:
\begin{quote}

---  The primary star has a mass
of $M_1=14\pm\,0.1M_\odot$ and a radius of $13.0\pm\,1.4R_\odot$.
The latter is nearly twice as large as that of a ZAMS star with the
same mass. The system orbit most likely has an inclination angle in
the range of $65^{\circ}-85^{\circ}$.

--- The  V/R variations for the H$_\alpha$ line of $\pi$ Aqr show a
sinusoidal behaviour with a period that coincides with the orbital
period of the system. Therefore, based on the results of
\citet{2002ApJ...573..812B}, \citet{2012IBVS.6023....1P}, and the
analysis presented here, we propose to include $\pi$ Aqr in the list
of Be binaries that show orbital phase-locked V/R variations. The
V/R variations reported here were coherent over  $\sim$40
cycles.

--- There is an S-wave in the $H_\alpha$  phased time series spectra, and the
Doppler tomograms demonstrate a corresponding bright extended spot
within the circumstellar disk. The radius of the circumstellar disk
around the primary component is $\sim65 R_\odot$.  The spot is
located in the outer part of the disk that faces the secondary. The
position of the spot in the disk is stable, but its relative
brightness correlates with the EW$_{H_\alpha}$ value. In general,
the structure of the spot looks like a one-armed spiral density
pattern. However, there is a faint hint of the second arm at the
opposite side of the disk. The brightest part of the spot begins at
$\approx -90^{\circ}$ from the major axis of the system and
continues to $\sim 120^{\circ}$ counterclockwise with decreasing
intensity (Figures~\ref{Dopmap} and \ref{Model}).
 \end{quote}

Long-term V/R variations in Be stars are well explained by a model
of global one-armed oscillations in the equatorial disk first
proposed by \citet{1983PASJ...35..249K} and developed by
\citet{1991PASJ...43...75O, 1997A&A...318..548O}. In this model the
one-armed perturbation slowly (on a time scale of $\sim10$ years)
precesses in the opposite direction to the semi-Keplerian motion as
a result of pressure forces in the disk. If we take into account
deviation from a $1/r$ point potential, this leads to a slowly
revolving prograde oscillation mode also with a very long time scale
semi-periods \citep{1992A&A...265L..45P}. The following studies
include other effects, although the number of free parameters and
their large range lead to a low predictive power of these models
\citep{2006A&A...447..277F}.

A significant number of Be stars are binary systems
\citep{2003PASP..115.1153P, 2010MNRAS.405.2439O,
2011IAUS..272..304M}. Nevertheless, in the current models, the
oscillation period weakly depends on the orbital parameters. For
example, \citet{2009PASJ...61...57O} found that the oscillation
period increases with increasing binary separation and/or decreasing
binary mass ratio. However, two well-established examples of orbital
phase-locked systems 4 Her and $\pi$~Aqr have relatively large
separations and small mass ratios of 0.06--0.016
\citep{1997A&A...328..551K} and 0.165 \citep{2002ApJ...573..812B},
respectively. Therefore, we must emphasize that, in fact, a number
of Be systems exist with phase-locked V/R variations.
Unfortunately, no explanation currently exists for this phenomenon
within the framework of the one-armed oscillation model. It is
likely that careful accounting of tidal or/and heating effects from
the secondary on the structure of a 
circumstellar disk can improve the situation.

There are also a number of Be + sdO binaries, such as 59\,Cyg
\citep{2005PAICz..93...21M, 2013ApJ...765....2P}, FY\,CMa
\citep{2005ASPC..337..309S, 2008ApJ...686.1280P} and $\phi$ Per
\citep{1993PASP..105..281G, 2000A&A...358..208S}), in which the
$V/R$ variations are also locked with the orbital period. The
orbital period locked emission in these systems is more likely
originated in the outer parts of the Be star disk facing the
secondary that are directly heated by a low mass ($\log M/M_\odot$
from $\sim -0.1$ to $\sim 0.1$), hot ($\log T_{\rm eff} \sim 4.7$)
subdwarf (sdO) companion \citep[see Table 5]{2013ApJ...765....2P}.
 Most sdO/sdB stars have masses between 0.40 $M_\odot$ and
0.55~$M_\odot$, but some of them may be as massive as $\sim
1.1M_\odot$ \citep{2010Ap&SS.329...11Z}. Although the effective
temperature of the secondary component in $\pi$~Aqr is unknown, its
mass $M_2$ can not be lower than 2 $M_\odot$ (Figure~\ref{fig5},
left panel). Therefore, it is unlikely to be an sdO or sdB star.

\citet{2002ApJ...573..812B} suggest that the secondary component in
$\pi$ Aqr is probably an A-- or F--type main-sequence star, since
only signs of the primary component are seen in the UV spectrum.
Nevertheless, the latter does not exclude that the mechanism of
phase-locked $V/R$ variations in $\pi$ Aqr can be similar or the
same (some tidal and heating effects on the structure of the
circumstellar disk caused by the secondary) as that discussed above
for Be + sdO binaries.

In any case, it is very important to continue observing $\pi$ Aqr
and other phase-locked systems spectroscopically, photometrically,
and interferometrically to search for more clues to them nature.
Finally, the data, presented in this paper, manifest a further
increasing role of the amateur spectroscopy in stellar astrophysics
\citep[cf.,][]{2013ApJ...766..119M}.

\begin{acknowledgements}
This work was supported  by DGAPA/PAPIIT project IN 103912. A.M.
acknowledges financial support from the University of North Carolina
at Greensboro and from its Department of Physics and Astronomy. We
thank observers at the Ritter Observatory for taking the spectra and
reducing them. This work has made use of the BeSS database, operated
at LESIA, Observatoire de Meudon, France: http://basebe.obspm.fr.

\end{acknowledgements}

\bibliographystyle{aa} 
\bibliography{szharikov}

\Online

\end{document}